\documentclass[preprint,nohyper]{JHEP3}

\usepackage{epsfig,multicol,latexsym}
\usepackage{amssymb,amsmath}
\usepackage{multirow,slashbox}
\usepackage{graphics,subfigure}
\usepackage{amsfonts}
\usepackage{multirow}
\usepackage{chngpage}

\newcommand{\dis}[1]{\begin{equation}\begin{split}#1\end{split}\end{equation}}

\newcommand{\be}{\begin{equation}}
\newcommand{\ee}{\end{equation}}

\newcommand{\gev}{\,\textrm{GeV}}
\newcommand{\mpl}{{m_\mathrm{Pl}}}
\newcommand{\ini}{{\mathrm{(in)}}}
\newcommand{\out}{{\mathrm{(out)}}}
\newcommand{\dec}{{\mathrm{(dec)}}}
\newcommand{\Gitot}{{\Gamma^{(i)}}}
\newcommand{\Girad}{{\Gamma_\gamma^{(i)}}}
\newcommand{\Gimat}{{\Gamma_m^{(i)}}}
\newcommand{\rhogzero}{{\rho_{\gamma 0}}}
\newcommand{\rhogi}{{\rho_{\gamma i}}}

\title{Multiple scalar particle decay \\ and perturbation generation}

\author{Ki-Young Choi \\
Department of Physics and Astronomy \\
University of Sheffield \\
Sheffield, S3 7RH \\
UK \\
\email{k.choi@sheffield.ac.uk}}

\author{Jinn-Ouk Gong \\
Harish-Chandra Research Institute \\
Chhatnag Road, Jhunsi\\
Allahabad 211 019 \\
India \\
\email{jgong@hri.res.in}}

\abstract{We study the evolution of the universe which contains a multiple number of
non-relativistic scalar fields decaying into both radiation and pressureless matter.
We present a powerful analytic formalism to calculate the matter and radiation
curvature perturbations, and find that our analytic estimates agree with full
numerical results within an error of less than one percent. Also we discuss the
isocurvature perturbation between matter and radiation components, which may be
detected by near future cosmological observations, and point out that it crucially
depends on the branching ratio of the decay rate of the scalar fields and that it is
hard to make any model independent predictions.}

\preprint{}

\begin{document}

\section{Introduction}
\label{intro}

Nowadays it is widely accepted that the primordial density perturbations are the
origin  of the temperature anisotropy in the cosmic microwave background (CMB) and
large scale structure in the present observable universe. Various cosmological
observations indicate that these perturbations are adiabatic and Gaussian, with an
almost scale invariant spectrum \cite{observations}. Interestingly, these
observational facts are consistent with an earlier inflationary era
\cite{inflation}: during inflation, quantum fluctuations of a slowly rolling scalar
field which dominates the energy density, the inflaton, are stretched and become
classical perturbations due to the quasi exponential expansion of the universe. A
particularly convenient quantity to study these perturbations is the curvature
perturbation $\zeta$ on uniform density hypersurfaces, developed
in~\cite{Bardeen:1983qw}, or $\mathcal{R}_c$ on comoving hypersurfaces which is
equivalent to $\zeta$ on large scales. For single field inflation cases $\zeta$ is
known to be conserved on large scales since perturbations are purely adiabatic, and
one can obtain the power spectrum of these perturbations with good enough accuracy
\cite{singlespectrum}. Note that, in multi-field inflationary models, in contrast,
there exists in general a non-adiabatic pressure perturbation and this makes $\zeta$
no more conserved on large scales\footnote{This is why the power spectrum of
primordial perturbations is evaluated only after the possible trajectories of the
inflaton fields coalesce in the so-called $\delta N$ formalism \cite{deltaN}.}
\cite{Wands:2000dp}.

In conventional inflationary models, the inflaton field is assumed to play two roles
at the same time: it dominates the energy density during inflation and makes the
universe expand enough to solve many cosmological problems such as homogeneity,
isotropy and flatness of the observable universe. Also, its vacuum fluctuations are
relevant for the curvature perturbation $\zeta$ and thus responsible for the
primordial density perturbations. Generally, the latter requirement introduce extra
fine tuning into the model: for example, in the simplest chaotic inflation model
with $V(\phi) = m^2\phi^2/2$, the inflaton mass $m$ can be as large as
$\mathcal{O}(\mpl)$, where $\mpl = (8\pi G)^{-1/2} \approx 2.4 \times 10^{18}\gev$
is the reduced Planck mass, when we do not mind perturbations and try to solve other
problems. However, to match the observed amplitude of density perturbations on large
scales, we need $m \sim \mathcal{O}\left(10^{-5}\right)\mpl$, i.e. we need a
relative fine tuning of one part over $10^5$ \cite{inflationbook}. However, during
inflation, {\em any} scalar fields with their masses being smaller than the Hubble
scale acquire almost scale invariant fluctuations. Such fields, depending on the
post-inflationary evolution of the universe, may later generate primordial density
perturbations by transferring their almost scale invariant isocurvature
perturbations to the curvature perturbation.

If this is the case, i.e. in the so-called curvaton scenario\footnote{There have
been some studies on similar scenario using the decay of neutrino dark matter
particles \cite{Khlopov:1999rs}.} \cite{curvaton}, such a field, dubbed the
``curvaton'', should satisfy several requirements: firstly, its effective mass must
be light, i.e. less than the Hubble parameter during inflation, to produce an almost
flat spectrum of fluctuations and to remain sub-dominant during inflation. It should
also couple very weakly to other fields so that its potential in the early universe
is not modified appreciably. It is also demanded that it keeps some level of
non-zero value \cite{nonzerocurvaton} and has not yet relaxed to its vacuum
expectation value. This is necessary to generate the appropriate amplitude of
perturbations. These conditions are basically what the conventional inflaton field
should satisfy, which is assumed to be responsible for the primordial density
perturbations, as well as the enough expansion of the universe. Thus, the curvaton
scenario may find its natural accommodation in the context of multi-field inflation
\cite{preparation}: for example, in a recently proposed scenario
\cite{Dimopoulos:2005ac} where a number of string axion fields drive inflation, it
is known \cite{Gong:2006zp} that there are a number of fields which have not yet
relaxed to their minima of the effective potential, with their mass being very small
relative to the Hubble parameter during inflation due to the assisted inflation
mechanism \cite{assistedinflation}.

Therefore, it is natural to consider the case where multiple curvaton fields are
responsible for the generation of the curvature perturbation after inflation. The
fluctuations of these curvaton fields are non-adiabatic in nature and thus, as
mentioned above, the curvature perturbation $\zeta$ does not remain constant but
evolves according to the energy transfer between different components which
constitute the universe. In this paper we study this general curvaton model. This
paper is outlined as follows. In Sec.~\ref{bgeqs}, we introduce the coupled
equations which determine the evolution of the universe. In Sec.~\ref{analytic} we
solve these equations analytically using the so-called sudden decay approximation,
using a novel and model independent method. In Sec.~\ref{apply} we apply our results
of the previous sections to several examples and compare the analytic estimates with
numerical calculations. Finally in Sec.~\ref{conclusions} we summarise and present
our conclusions.

\section{Background equations and perturbations}
\label{bgeqs}

In this section we will summarise the evolution of the background quantities in a
flat universe and show the evolution equations of the curvature perturbations of the
components in the system of multiple curvatons decaying into radiation and matter.
We assume that the universe is initially dominated by radiation due to the decay of
the inflaton field(s) after inflation.

We assume that the curvatons ($\sigma_i$) decay into both radiation ($\gamma$) and
non-relativistic matter ($m$) with constant decay rates $\Girad$ and $\Gimat$
respectively, which are fixed by underlying physics. The energy transfer equations
between components are given by\footnote{One can add the effect of dark matter
freeze-out and annihilation\cite{Lemoine:2006sc}, but the qualitative evolution is
not too different.}
\begin{align}\label{Qi}
Q_i & = -(\Girad + \Gimat)\rho_i \equiv -\Gitot\rho_i \, ,
\\\label{Qg}
Q_\gamma & = \sum_i \Girad\rho_i \equiv \sum_i Q_{\gamma i} \, ,
\\\label{Qm}
Q_m & = \sum_i \Gimat\rho_i \equiv \sum_i Q_{mi} \, ,
\end{align}
where we have introduced the total decay width of $\sigma_i$,
$\Gitot \equiv \Girad + \Gimat$, and the energy transfer to radiation (matter)
by the decay of $\sigma_i$, $Q_{\gamma i}$ ($Q_{mi}$), respectively.
Note that they obey the constraint of energy conservation
\dis{ \sum_iQ_i+Q_\gamma+ Q_m =0 \, . }
Thus from the general continuity equation of each component including energy
transfer~\cite{Kodama:1985bj},
\begin{equation}
\dot\rho_\alpha = -3H(\rho_\alpha + p_\alpha) + Q_\alpha \, ,\label{continuity_Eq}
\end{equation}
we find that for each component
\begin{align}\label{continuity_Eq_i}
\dot\rho_i & = - (3H + \Gamma^{(i)})\rho_i \, ,
\\\label{continuity_Eq_gamma}
\dot\rho_\gamma & = -4H\rho_\gamma + \sum_i \Gamma_\gamma^{(i)}\rho_i \, ,
\\\label{continuity_Eq_m}
\dot\rho_m & = -3H\rho_m + \sum_i \Gamma_m^{(i)}\rho_i \, .
\end{align}
Note that we can obtain the continuity equation of the total energy density by
summing over that of each component,
\begin{align}
\dot\rho & = -3H(\rho + p)
\nonumber\\
& = -H \left( 4\rho_\gamma + 3\rho_m + 3\sum_i \rho_i \right) \, ,
\end{align}
where the total density $\rho$ and pressure $p$ are given by
\begin{align}\label{totalrho}
\rho & = \rho_\gamma +\rho_m + \sum_i\rho_i \, ,
\\
p & = p_\gamma + p_m + \sum_ip_i \, ,
\end{align}
respectively. In the above we take $p_\gamma = \rho_\gamma/3$ and $p_m =
p_i = 0$, i.e. the equation of state of the curvaton fields are effectively
equivalent to that of pressureless matter.

By adopting the density parameters $\Omega_\gamma$, $\Omega_m$ and $\Omega_i$, we
can rewrite Eqs.~(\ref{continuity_Eq_i}), (\ref{continuity_Eq_gamma}) and
(\ref{continuity_Eq_m}) in more convenient dimensionless forms for numerical
calculation. From the Friedmann equation
\begin{equation}\label{friedmann}
H^2 = \frac{1}{3m_\mathrm{Pl}^2}\rho \, ,
\end{equation}
the density parameters satisfy the relation
\begin{equation}
\Omega_\gamma + \Omega_m + \sum_i \Omega_i = 1 \, .
\end{equation}
Then, Eqs.~(\ref{continuity_Eq_i}), (\ref{continuity_Eq_gamma}) and
(\ref{continuity_Eq_m}) can be written as
\begin{align}\label{Oieq}
\Omega_i' & = \Omega_i \left( \Omega_\gamma - H^{-1}\Gitot \right) \, ,
\\\label{Ogeq}
\Omega_\gamma' & = H^{-1} \sum_i \Omega_i\Girad - \Omega_\gamma (1 - \Omega_\gamma)
\, ,
\\\label{Omeq}
\Omega_m' & = H^{-1} \sum_i \Omega_i\Gimat + \Omega_\gamma\Omega_m \, ,
\end{align}
and Eq.~(\ref{friedmann}) as
\begin{equation}\label{friedmannOg}
H' = -\frac{3 + \Omega_\gamma}{2} H \, ,
\end{equation}
where a prime denotes a derivative with respect to the number of $e$-folds,
\begin{equation}
N \equiv \int H dt \, .
\end{equation}

The total curvature perturbation on uniform curvature hypersurfaces is given by
\begin{equation}
\zeta = -H \frac{\delta\rho}{\dot\rho} \, ,
\end{equation}
which can be written as a weighted sum of the curvature perturbation of the
component $\alpha$ on the corresponding uniform density hypersurfaces
$\zeta_\alpha$
\cite{Wands:2000dp},
\begin{equation}
\zeta = \sum_\alpha \frac{\dot\rho_\alpha}{\dot\rho}\zeta_\alpha \, ,
\label{zeta_overall}
\end{equation}
where
\begin{equation}\label{zetacomponent}
\zeta_\alpha = -H \frac{\delta\rho_\alpha}{\dot\rho_\alpha} \, .
\end{equation}
The difference between any two components gives an isocurvature perturbation~\cite{Malik:2002jb}
\begin{equation}
\mathcal{S}_{\alpha\beta} = 3(\zeta_\alpha - \zeta_\beta) \, .
\end{equation}
The total curvature perturbation on large scales evolves as \cite{Wands:2000dp}
\begin{equation}\label{zetaevolution}
\dot\zeta = -\frac{H}{\rho + p}\delta p_\mathrm{nad} \, ,
\end{equation}
where the non-adiabatic pressure perturbation is given by
\begin{equation}
\delta p_\mathrm{nad} \equiv \delta p - \frac{\dot{p}}{\dot\rho}\delta\rho \, .
\end{equation}
Therefore, as mentioned before, $\zeta$ remains constant on large scales when the
perturbations are purely adiabatic. From Eqs.~(\ref{zetacomponent}) and
(\ref{zetaevolution}), and using the perturbed continuity equations of each
component~\cite{Malik:2002jb,Gupta:2003jc}, we can find that the curvature
perturbations of the components evolve on large scales as
\begin{align}\label{zieq}
\zeta_i' & = \frac{H^{-1}\Gitot (3 + \Omega_\gamma)}{2 \left( 3 + H^{-1}\Gitot
\right)} \left( \zeta - \zeta_i \right) \, ,
\\\label{zgeq}
\zeta_\gamma' & = \left( 4\Omega_\gamma - H^{-1}\sum_i\Girad\Omega_i \right)^{-1}
\left[ \sum_j \Omega_j (3 + H^{-1}\Gitot) H^{-1}\Girad \left( \zeta_i - \zeta_\gamma
\right) \right.
\nonumber\\
& \hspace{5.5cm} \left. - \frac{H^{-1}\sum_k\Gamma_\gamma^{(k)}\Omega_k}{2} (3 +
\Omega_\gamma) \left( \zeta - \zeta_\gamma \right) \right] \, ,
\\\label{zeq}
\zeta' & = \frac{4\Omega_\gamma - H^{-1}\sum_i \Girad\Omega_i}{3 + \Omega_\gamma}
\left( \zeta - \zeta_\gamma \right) \, .
\end{align}
Here we do not solve the evolution of $\zeta_m$ directly, though it is
straightforward to write the evolution equation of $\zeta_m$: rather, from
Eq.~(\ref{zeta_overall}), $\zeta_m$ is calculated as
\begin{equation}
\zeta_m = \frac{\dot\rho\zeta - \dot\rho_\gamma\zeta_\gamma -
\sum_i\dot\rho_i\zeta_i}{\dot\rho_m} \, .
\end{equation}
The reason is the existence of singularity in $\zeta_m$, because there exists some
moment $\dot\rho_m = 0$ when the dilution of matter due to the expansion of the
universe is balanced with the creation of matter due to the curvaton
decay\footnote{In fact this is the same for radiation component. However, as long as
we assume that the density of radiation is initially high so that the universe is
radiation dominated, $\dot\rho_\gamma < 0$ always.}
\cite{Malik:2002jb,Gupta:2003jc}.

We may solve Eqs.~(\ref{Oieq})--(\ref{friedmannOg}) and (\ref{zieq})--(\ref{zeq})
numerically, which would be the simplest way to study the evolution of the curvature
perturbation. However, we can obtain further insights by implementing analytic
analysis. In the following section we will find the final curvature perturbations
under the so-called sudden decay approximation \cite{Lyth:2002my}.

\section{Analytic approximation }
\label{analytic}

In this section, we study the curvature perturbations under the assumption that
there is no interaction between components until the curvaton fields decay and that
the decay of each curvaton is instantaneous. Under this ``sudden decay
approximation'', we can derive analytic estimates for the curvature perturbations
associated with matter and radiation after all the curvaton fields decay, as we will
see in this section. Note that from Eqs.~(\ref{Qi}), (\ref{Qg}) and (\ref{Qm}),
after all the curvatons decay, there is no energy transfer between matter and
radiation and hence $\zeta_m$ and $\zeta_\gamma$ are constant, though $\zeta$ will
still evolve on large scales. In this sense, we will call these $\zeta_m$ and
$\zeta_\gamma$ after the decay of the curvatons as ``final'' curvature
perturbations, and denote by the superscript (out).

For our purpose in this section, we decompose the radiation and matter density
according to the source of generation,
\dis{
\rho_\gamma &= \rhogzero + \sum_{i}\rhogi,\\
\rho_m & = \rho_{m 0} +\sum_i \rho_{m i},\label{split_rho} }
where $\rhogzero$ ($\rho_{m0}$) is the energy density of radiation (matter) which is
due to the decay of the inflaton field(s) and independent of the curvaton decay, and
$\rhogi$ ($\rho_{m i}$) is the radiation (matter) density generated from the decay
of $\sigma_i$ \cite{Hamaguchi:2003dc}. Then, Eq.~(\ref{totalrho}) can be written as
\begin{align}\label{compositedensity}
\rho & = \rhogzero + \rho_{m0} + \sum_i \left( \rhogi + \frac{\Girad}{\Gitot}\rho_i
\right) + \sum_i \left( \rho_{mi} + \frac{\Gimat}{\Gitot}\rho_i \right)
\nonumber\\
& \equiv \rhogzero + \rho_{m0} + \sum_i \widetilde\rho_{\gamma i} + \sum_i
\widetilde\rho_{mi} \, ,
\end{align}
where we have introduced two composite densities $\widetilde\rho_{\gamma i}$ and
$\widetilde\rho_{mi}$ which will play the central role in the discussions below.

\subsection{Matter curvature perturbation}

From Eqs.~(\ref{Qi}), (\ref{Qm}) and (\ref{compositedensity}), we can see that for
the composite density $\widetilde\rho_{mi}$,
\begin{equation}
\widetilde{Q}_{mi} = Q_{mi} + \frac{\Gimat}{\Gitot}Q_i = 0 \, ,
\end{equation}
i.e. the energy transfer is zero. Moreover, since the corresponding equation of
state is that of pressureless matter, we can write
\begin{equation}
\dot{\widetilde\rho}_{mi} = -3H\widetilde\rho_{mi} \, ,
\end{equation}
and therefore the associated curvature perturbation \cite{Gupta:2003jc},
\begin{equation}
\widetilde\zeta_{mi} = -H \frac{\delta\widetilde\rho_{mi}}{\dot{\widetilde\rho}_{mi}} =
\frac{\delta\widetilde\rho_{mi}}{3\widetilde\rho_{mi}} \, ,
\end{equation}
is conserved on large scales. Well before the curvaton $\sigma_i$ decays $\rho_{mi}
= \delta\rho_{mi} = 0$ so $\widetilde\zeta_{mi} = \zeta_i^\ini$, meanwhile after
$\sigma_i$ decays $\rho_i$ is negligible and thus $\widetilde\zeta_{mi} =
\zeta_{mi}^\out$. Therefore we have
\begin{equation}\label{composite_zm}
\zeta_{mi}^\out = \zeta_i^\ini \, .
\end{equation}
Thus, from Eqs.~(\ref{zetacomponent}), (\ref{split_rho}) and (\ref{composite_zm}),
we find that the final matter curvature perturbation after all the curvatons decay
is given by
\begin{align}\label{zeta_m_out}
\zeta_m^\out & = \left( \rho_{m0}^\ini \zeta_{m0}^\ini + \sum_i
\frac{\Gimat}{\Gitot} \rho_i^\ini \zeta_i^\ini \right) \left[ \rho_{m0}^\ini +
\sum_j \frac{\Gamma_m^{(j)}}{\Gamma^{(j)}} \rho_j^\ini \right]^{-1}
\nonumber\\
& \equiv \sum_{i = 0}^n s_i \zeta_i^\ini \, ,
\end{align}
where $\zeta_0^\ini \equiv \zeta_{m0}^\ini$. The transfer coefficient $s_i$ we have
introduced above is given by
\begin{align}\label{s_i}
s_0 & = \Omega_{m0}^\ini \left[ \Omega_{m0}^\ini + \sum_j
\frac{\Gamma_m^{(j)}}{\Gamma^{(j)}} \Omega_j^\ini \right]^{-1} \, ,
\nonumber\\
s_i & = \frac{\Gimat}{\Gitot} \Omega_i^\ini \left[ \Omega_{m0}^\ini + \sum_j
\frac{\Gamma_m^{(j)}}{\Gamma^{(j)}} \Omega_j^\ini \right]^{-1} \, . \hspace{.5cm} (i
= 1, 2, \cdots, n)
\end{align}
So we can see that the final matter curvature perturbation is completely determined
by the decay rate and the initial energy density $\rho_i^\ini$, or equivalently,
initial density parameter $\Omega_i^\ini$ of each curvaton field and that of
pre-existing matter as shown above.

\subsection{Radiation curvature perturbation}

In the previous section, we could use the conservation of the curvature perturbation of each composite density $\widetilde\rho_{mi}$ to
find out the final matter curvature perturbation. This is possible since every $\widetilde{\rho}_{mi}$ has no
energy transfer and in addition unique equation of state. One may hope that similar argument is
applied to the other composite density we have introduced in
Eq.~(\ref{compositedensity}), $\widetilde\rho_{\gamma i}$, but this is not the case.
Nevertheless, $\widetilde\rho_{\gamma i}$ turns out to be an useful quantity to
calculate the final radiation curvature perturbation as we will see shortly. In this
section, we assume that the decay rates of the curvaton fields are different so that they do not decay at the same time: rather, they decay successively due
to different decay rates. Without loss of generality we put the order of
curvatons by the decay rate of each curvaton to satisfy $\Gitot>\Gamma^{(i+1)}$.

First we consider a limited time interval around the decay of the curvaton field
$\sigma_1$, which is assumed to have the largest decay rate. We write a combined
density of radiation and the curvaton field
\begin{align}
\rho_\gamma^{(1)} & \equiv \rhogzero + \widetilde\rho_{\gamma 1}
\nonumber\\
& = \rhogzero + \rho_{\gamma 1} + \frac{\Gamma_\gamma^{(1)}}{\Gamma^{(1)}} \rho_1 \,
.
\end{align}
Note that although the energy transfer of $\rho_\gamma^{(1)}$ is zero, its equation
of state is not unique and thus the corresponding curvature perturbation
$\zeta_\gamma^{(1)}$ evolves on large scales. Therefore, as mentioned above, unlike
$\widetilde\rho_{mi}$ we cannot simply connect the initial curvature perturbations in
the curvaton fields to the final one in radiation, but rather we have to get through
the moments of decay. Now we assume that until $\sigma_1$ decays instantaneously
there is no energy transfer between the curvaton and radiation. Then,
$\rho_\gamma^{(1)}$ before and after $\sigma_1$ decays, which we write respectively
\begin{align}
\rho_\gamma^{(1)}|_\mathrm{before} & = \rhogzero^{<1} +
\frac{\Gamma_\gamma^{(1)}}{\Gamma^{(1)}} \rho_1^{<1} \, ,
\\
\rho_\gamma^{(1)}|_\mathrm{after} & = \rhogzero^{>1} + \rho_{\gamma 1}^{>1} \, ,
\end{align}
where the superscript $<1$ ($>1$) means that it is evaluated before (after)
$\sigma_1$ decays, and these densities have the same value at the moment of decay. Since $\rho_{\gamma
1}$ is generated only after $\sigma_1$ decays, the value of $\rho_{\gamma 1}$ at the
moment of decay corresponds to
its initial value and thus
\begin{equation}
\rho_{\gamma 1}^\dec = \frac{\Gamma_\gamma^{(1)}}{\Gamma^{(1)}}\rho_1^\dec \, .
\end{equation}
Using the fact that both $\rhogzero$ and $\rho_{\gamma 1}$ scale as $a^{-4}$, we can
write the ratio $\rho_{\gamma 1}/\rhogzero$ at late times, which is constant after
$\sigma_1$ decays, as
\begin{equation}\label{rho1rho0ratio}
\frac{\rho_{\gamma 1}}{\rhogzero} = \frac{\rho_{\gamma 1}^\dec
(a^\dec/a)^4}{\rhogzero^\dec(a^\dec/a)^4} = \frac{\Gamma_\gamma^{(1)}}{\Gamma^{(1)}}
\frac{\rho_1^\dec}{\rhogzero^\dec} \, .
\end{equation}
The individual curvature perturbations $\zeta_{\gamma 0}$ and $\zeta_1$ remain
constant on large scales before $\sigma_1$ decays. Then, the combined curvature
perturbation $\zeta_\gamma^{(1)}$ corresponding to $\rho_\gamma^{(1)}$ is written as
\begin{equation}
\zeta_\gamma^{(1)} \approx (1 - f_1)\zeta_{\gamma 0} + f_1\zeta_1 \, ,
\end{equation}
where
\begin{equation}\label{f1}
f_1 = \frac{3\frac{\Gamma_\gamma^{(1)}}{\Gamma^{(1)}}\rho_1}{4\rhogzero + 3
\frac{\Gamma_\gamma^{(1)}}{\Gamma^{(1)}}\rho_1} \, .
\end{equation}
Here $f_1$, the weight of $\zeta_1$, solely describes the evolution of $\zeta_\gamma^{(1)}$ on
large scales. After the curvaton $\sigma_1$ decays, the energy density
$\rho_\gamma^{(1)}$ is identical to $\rho_\gamma$ at that time and has a unique
equation of state. Hence after the decay of $\sigma_1$,
$\zeta_\gamma^{(1)}$ becomes constant on large scales until the curvaton with the
next largest decay width begins to decay, i.e.~\cite{Malik:2002jb}
\begin{equation}\label{zetagammaafter1}
\zeta_\gamma^{>1} = \zeta_\gamma^{(1)}|_\mathrm{dec} \approx \left( 1 - f_1^\dec
\right) \zeta_{\gamma 0}^\ini + f_1^\dec \zeta_1^\ini \, ,
\end{equation}
where, using Eq.~(\ref{rho1rho0ratio}), $f_1^\dec$ is given by
\begin{equation}
f_1^\dec = \frac{3\rho_{\gamma 1}/\rhogzero}{4 + 3 \rho_{\gamma 1}/\rhogzero} \, .
\end{equation}

We can take the same step for the successive curvaton decays: e.g. for $\sigma_2$
which has the next largest decay width, we just replace
\begin{align}
\rho_\gamma^{(1)}|_\mathrm{after} & \Rightarrow \rho_{\gamma 0 \mathrm{(new)}} \, ,
\nonumber\\
\zeta_\gamma^{>1} & \Rightarrow \zeta_{\gamma 0 \mathrm{(new)}} \, ,
\end{align}
and so on. In general after $i$-th curvaton $\sigma_i$ decays, the curvature
perturbation in the radiation component is constant until the decay of $(i+1)$-th
curvaton, and is written as
\begin{equation}
\zeta_\gamma^{>i} \approx \left( 1 - f_i^\dec \right) \zeta_\gamma^{>i-1} +
f_i^\dec\zeta_i^\ini \, ,
\end{equation}
where
\begin{equation}
f_i^\dec = \frac{3\rhogi/\rhogzero}{4\sum_{k = 0}^{i-1} \rho_{\gamma k}/\rhogzero +
3\rhogi/\rhogzero} \, . \label{f_i_dec}
\end{equation}
Therefore, after all the $n$ curvatons decay, we find the final curvature
perturbations in radiation as
\begin{align}
\zeta_\gamma^\out & \approx \left( 1 - f_n^\dec \right) \zeta_\gamma^{>n-1} +
f_n^\dec\zeta_n^\ini
\nonumber\\
& = \left( 1 - f_n^\dec \right) \left( 1 - f_{n-1}^\dec \right) \zeta_\gamma^{>n-2}
+ \left( 1 - f_n^\dec \right) f_{n-1}^\dec \zeta_{n-1}^\ini + f_n^\dec\zeta_n^\ini
\nonumber\\
& = \cdots
\nonumber\\
& \equiv \sum_{i=0}^n r_i \zeta_i^\ini \, , \label{zeta_gamma_out}
\end{align}
where $\zeta_0^\ini \equiv \zeta_{\gamma 0}^\ini$ and $f_0^\dec = 1$.
The transfer coefficient $r_i$ is given by
\begin{align}
r_i & = \prod_{k = i + 1}^n \left(1 - f_k^\dec \right) f_i^\dec
=\left(1-\sum_{k=i+1}^{n}r_k\right) f_i^\dec \, , \hspace{.5cm}
(i = 0,1,\cdots,n - 1)
\nonumber\\
r_n & = f_n^\dec\, ,\label{r_i}
\end{align}
and is completely determined once we find the ratio $\rhogi/\rhogzero$.

\subsection{Ratio of radiation after curvaton decay}

We found in the previous section that the final radiation curvature perturbation
depends on the ratio of the radiation generated from curvaton decay with respect to
the original radiation component. In this section, we present a general and simple
way to calculate this ratio analytically.

\begin{figure}[!t]
  \begin{center}
  \begin{tabular}{c}
    \includegraphics[width=0.45\textwidth]{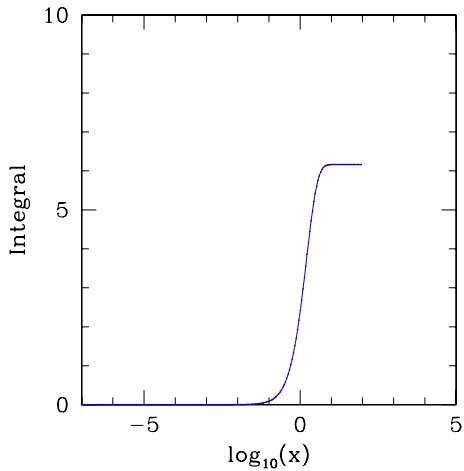}
    \end{tabular}
  \end{center}
\caption{Plot of the integral $\int_0^{x_i}y \exp(-u_i) du_i$ versus $x\equiv x_1$.
The value changes quickly only near the moment of decay time, $t\sim \Gamma^{-1}$, and
becomes constant afterwards.}
\label{fig:integral}
\end{figure}

From Eqs.~(\ref{continuity_Eq_i}), (\ref{continuity_Eq_gamma}) and
(\ref{continuity_Eq_m}), we can write the continuity equations of the components
used in Eq.~(\ref{compositedensity}) as
\begin{align}
\dot\rho_{\gamma 0} & = -4H\rhogzero \, ,
\\
\dot\rho_{m0} & = -3H\rho_{m0} \, ,
\\
\dot{\widetilde\rho}_{\gamma i} & = -4H\rhogi - 3H \frac{\Girad}{\Gitot} \rho_i \, ,
\\
\dot{\widetilde\rho}_{mi} & = -3H\rho_{mi} - 3H \frac{\Gimat}{\Gitot} \rho_i \, .
\end{align}
We can solve these equation analytically and the solutions are given by
\begin{align}
\rhogzero & = \rho_{\gamma 0}^\ini \left( \frac{a^\ini}{a} \right)^4 \, ,
\\
\rho_{m0} & = \rho_{m0}^\ini \left( \frac{a^\ini}{a} \right)^3 \, ,
\\
\rho_i & = \rho_i^\ini \left( \frac{a^\ini}{a} \right)^3 \exp \left[ -\Gitot (t -
t_0) \right] \, ,
\\
\rhogi & = \Girad\rho_i^\ini \left( \frac{a^\ini}{a} \right)^4 \int_{t_0}^t
\frac{a}{a^\ini} \exp \left[ -\Gitot t' \right] dt' \, ,
\\
\rho_{mi} & = \frac{\Gimat}{\Gitot} \rho_i^\ini \left( \frac{a^\ini}{a} \right)^3
\left\{ 1 - \exp \left[ -\Gitot (t - t_0) \right] \right\} \, ,
\end{align}
where we have set the initial time to be $t_0$. Now, introducing
\cite{Scherrer:1984fd}
\begin{align}
z  \equiv \frac{a}{a^\ini} \, , \qquad x_i  \equiv \Gitot t \, ,
\end{align}
and using Eq.~(\ref{compositedensity}), the Friedmann equation,
\begin{equation}
H^2 = \frac{1}{3m_\mathrm{Pl}^2} \left[ \rhogzero + \rho_{m0} + \sum_i \left( \rhogi
+ \rho_{mi} + \rho_i \right) \right] \, ,
\end{equation}
becomes
\begin{equation}
\left( \frac{z'}{z} \right)^2 = x_H^{-2} \left\{ z^{-4} +
\frac{\Omega_{m0}^\ini}{\Omega_{\gamma 0}^\ini}z^{-3} + \sum_i \left[
\frac{\Girad}{\Gitot} \frac{\Omega_i^\ini}{\Omega_{\gamma 0}^\ini} z^{-3} e^{-x_i} +
\frac{\Gimat}{\Gitot} \frac{\Omega_i^\ini}{\Omega_{\gamma 0}^\ini} z^{-3} +
\frac{\Girad}{\Gitot}\frac{\Omega_i^\ini}{\Omega_{\gamma 0}^\ini} z^{-4}
\int_{x_0}^{x_i} z e^{-u_i} du_i \right] \right\} \, ,
\end{equation}
where
\begin{equation}\label{xH}
x_H \equiv {\Omega_{\gamma 0}^\ini}^{-1/2} \frac{\Gamma^{(1)}}{H^\ini} \, ,
\end{equation}
$u_i \equiv \Gitot t'$ and a prime denotes a derivative with respect to $x_1$. We
choose $x_1$ for convenience since the dependence on this particular choice of $x_1$
is absorbed into the definition of $x_H$ as shown above. Finally, introducing a new
variable
\begin{equation}
y \equiv x_H^{1/2} z \, ,
\end{equation}
we finally obtain the dimensionless Friedmann equation
\begin{equation}\label{modifiedFriedmann}
\left( \frac{y'}{y} \right)^2 = y^{-4} +\beta_0 y^{-3} + \sum_i \left[ \alpha_i y^{-3}
e^{-x_i} + \beta_i y^{-3} + \alpha_i y^{-4} \int_{x_0}^{x_i} y e^{-u_i} du_i \right]
\, ,
\end{equation}
where the coefficients $\alpha_i$, $\beta_i$ and $\beta_0$ are defined by
\dis{
\alpha_i\equiv \frac{\Girad}{\Gitot} \frac{\Omega_i^{\ini}}{\Omega_{\gamma
0}^{\ini}}x_H^{-1/2},\qquad
\beta_i\equiv \frac{\Gimat}{\Gitot}
\frac{\Omega_i^{\ini}}{\Omega_{\gamma 0}^{\ini}}x_H^{-1/2}, \qquad
\beta_0\equiv \frac{\Omega_{m0}^\ini}{\Omega_{\gamma 0}^\ini} x_H^{-1/2} \, , \label{ratiocoeff}}
respectively. Then the ratio of radiations which determines the transfer coefficient
$r_i$ is given by
\begin{equation}\label{rhogammaratio}
\frac{\rhogi}{\rhogzero} = \alpha_i \int_{x_0}^{x_i} ye^{-u_i} du_i \, ,
\end{equation}
where the integrand is suppressed exponentially after the curvaton $\sigma_i$
decays, and the integral becomes almost constant. In Fig.~\ref{fig:integral}, we
plot this integral as a function of $x\equiv \Gamma t$. A large change occurs only
around the decay time ($x\sim 1$) and soon becomes constant. We can see that the
most significant contribution of this integral comes from the epoch around the
moment of decay.

\subsection{Final curvature and isocurvature perturbations}

After all the curvaton fields decay, i.e. $\Omega_i = 0$, we are left with the
overall curvature perturbation given by, from Eq.~(\ref{zeta_overall}),
\begin{align}\label{finalzeta}
\zeta & = \frac{\dot\rho_\gamma}{\dot\rho}\zeta_\gamma^\out +
\frac{\dot\rho_m}{\dot\rho}\zeta_m^\out
\nonumber\\
& = \frac{4\Omega_\gamma}{4\Omega_\gamma + 3\Omega_m}\zeta_\gamma^\out +
\frac{3\Omega_m}{4\Omega_\gamma + 3\Omega_m}\zeta_m^\out \, .
\end{align}
The final matter and radiation curvature perturbations are constant on large scales
and given by Eqs.~(\ref{zeta_m_out}) and~(\ref{zeta_gamma_out}), respectively. Their
transfer coefficients are determined by Eqs.~(\ref{s_i}), (\ref{r_i}) and
(\ref{rhogammaratio}). Thus, the isocurvature perturbation between matter and
radiation components $\mathcal{S}_{m\gamma} = 3(\zeta_m - \zeta_\gamma)$, which is
fixed after all the curvaton fields decay so that $\zeta_\gamma$ and $\zeta_m$
become constants, is written as
\begin{align}\label{finaliso}
\mathcal{S}_{m\gamma}^\out & = 3 \left( \zeta_m^\out - \zeta_\gamma^\out \right)
\nonumber\\
& = 3 \sum_i (s_i - r_i) \zeta_i^\ini \, .
\end{align}

A particularly simple case is when all the decay rates are the same:
then, from Eq.~(\ref{s_i}), the transfer coefficient of matter curvature
perturbation becomes simply
\begin{equation}
s_i = \frac{\Omega_i^\ini}{\sum_j\Omega_j^\ini} = \frac{\Omega_i^\ini}{1 -
\Omega_\gamma^\ini} \, ,
\end{equation}
where we have assumed that initially there is no matter component. As can be seen
clearly, the most significant contribution to the final matter curvature
perturbation comes from the curvaton field which initially occupies the largest
energy density among the curvatons. For $r_i$, we only need to consider a single
moment of decay since the curvaton fields decay at the same time. Thus, from
Eqs.~(\ref{rho1rho0ratio}) and (\ref{f1}), we simply have
\begin{equation}
f_i^\dec = \frac{3\rhogi/\rhogzero}{4 + 3\sum_j\rho_{\gamma j}/\rhogzero} \, ,
\end{equation}
and the final radiation curvature perturbation becomes, from
Eq.~(\ref{zetagammaafter1}),
\begin{equation}
\zeta_\gamma^\out = \left( 1 - \sum_i f_i^\dec \right) \zeta_{\gamma 0}^\ini +
\sum_i f_i^\dec \zeta_i^\ini \, .
\end{equation}
Now, from Eqs.~(\ref{ratiocoeff}) and (\ref{rhogammaratio}), we can see that the
ratio $\rhogi/\rhogzero$ is proportional to $\alpha_i$, which is again proportional
to $\Omega_i^\ini$, since [integral] $\equiv \int_{x_0}^{x_i} y\exp(-u_i)du_i$ will
 have the same value as discussed in the previous section. Hence,
\begin{equation}
r_i=f_i^\dec = \frac{3\mathrm{[integral]}\alpha_i}{4 +
3\mathrm{[integral]}\sum_j\alpha_j} = \frac{3C\mathrm{[integral]}\Omega_i^\ini}{4 +
3C\mathrm{[integral]}\sum_j\Omega_j^\ini} \, ,
\end{equation}
where $C = \Gamma_\gamma
{H^\ini}^{1/2}/\left(\Gamma^{3/2}{\Omega_\gamma^\ini}^{3/4}\right)$ is the common
coefficient of proportionality of $\alpha_i$ to $\Omega_i^\ini$. Thus, with one
further assumption that the initial radiation curvature perturbation is negligible,
i.e. $\zeta_{\gamma 0}^\ini \approx 0$, the final isocurvature perturbation is, from
Eq.~(\ref{finaliso}),
\begin{equation}
\mathcal{S}_{m\gamma}^\out \approx 3\sum_i \left[ \frac{1}{1 - \Omega_\gamma^\ini} -
\frac{3C\mathrm{[integral]}}{4 + 3C\mathrm{[integral]}\left( 1 - \Omega_\gamma^\ini
\right)} \right] \Omega_i^\ini \zeta_i^\ini \, ,
\end{equation}
and thus the transfer from the initial curvature perturbation $\zeta_i^\ini$ is
proportional to the corresponding initial density fraction $\Omega_i^\ini$.

\section{Applications}
\label{apply}

In this section, we apply our analytic estimates obtained in the previous section to
several examples and compare with numerical results.

\subsection{Single curvaton}

First we consider a simple example where a single curvaton field decays into
radiation and matter with decay rates $\Gamma^{(1)}_\gamma$ and $\Gamma^{(1)}_m$,
respectively. If we assume that the initial curvature perturbation in radiation is
negligible, which is usually taken as the initial condition for the curvaton
scenario, the radiation curvature perturbation after curvaton decay is purely due to
the decay of the curvaton field and from Eqs.~(\ref{zeta_gamma_out}) is given by
\dis{ \zeta_\gamma^\out \approx f_{1}^\dec \zeta_1^\ini \, , \label{zeta1_out} }
where
\dis{ f_{1}^\dec= \frac{3\rho_{\gamma 1}/\rhogzero}{4+3\rho_{\gamma 1}/\rhogzero} \,
. \label{f_1_dec} }
As discussed in the previous sections this is constant after the curvaton decay, and
is completely determined once we find the ratio $\rho_{\gamma 1}/\rhogzero$. This
ratio is given by Eq.~(\ref{rhogammaratio}) as
\dis{ \frac{\rho_{\gamma 1}}{\rhogzero}&=\alpha_1\int_{x_0}^{x_1}y(u_1)e^{-u_1}du_1
\, ,\label{ratio_1} }
and depends only on $x_H$ and $\alpha_1$ given by Eqs.~(\ref{xH}) and
(\ref{ratiocoeff}), respectively.

If initially radiation dominates, i.e. $\Omega_{\gamma 0}^\ini \approx 1$, we find
that
\begin{align}
x_H & \approx \frac{\Gamma^{(1)}}{H^\ini} \, ,
\label{xHsingle}\\
\alpha_1 & \approx \Omega_1^\ini \frac{\Gamma_\gamma^{(1)}}{\Gamma^{(1)}} \left(
\frac{\Gamma^{(1)}}{H^\ini} \right)^{-1/2}
\label{alpha1single}
\end{align}
where $\alpha_1$ becomes identical with $p$ of Ref.~\cite{Gupta:2003jc} in the limit
$\Gamma_\gamma^{(1)} \gg \Gamma_m^{(1)}$. In this case since the universe is
dominated by radiation component, $a \propto t^{1/2}$ and $H = (2t)^{-1}$. That is,
for small $x_1$ the solution of Eq.~(\ref{modifiedFriedmann}) is given by
\dis{ y(x_1)= (2x_1)^{1/2} \, , }
with $x_1^\ini=x_H/2$, and we can see that $y(x_1)$ is independent of
$x_H$~\cite{Scherrer:1984fd}. Thus the curvature perturbation depends only on
$\alpha_1$, which is shown by using the phase space plot in
Refs.~\cite{Malik:2002jb,Gupta:2003jc}.

Furthermore, in the case that the curvaton does not dominate the density during the
evolution, we can further approximate Eq.~(\ref{ratio_1}) analytically. From the
sudden decay approximation, we can see that $\alpha_1+\beta_1 \approx\Omega_1^\ini
(\Gamma^{(1)}/H^\ini)^{-1/2} \ll 1$ guarantees the radiation
domination~\cite{Malik:2002jb}, thus using $y(t)\simeq x_H^{1/2}(t/t_0)^{1/2}$, we
obtain
\begin{align}
\frac{\rho_{\gamma 1}}{\rhogzero} & \approx \alpha_1 x_H^{1/2} \int_{x_0}^{x_1}
\left( \frac{t}{t_0} \right)^{1/2} e^{-u_1}du_1
\nonumber\\
& = \frac{\alpha_1x_H^{1/2}}{\sqrt{t_0\Gamma^{(1)}}}
\int_{x_0}^{x_1} u_1^{1/2}e^{-u_1} du_1
\nonumber\\
& \approx \sqrt{2}\alpha_1 \int_{x_0}^{x_1} u_1^{1/2} e^{-u_1} du_1 \, ,
\label{rhogamma1ratio-ex1}
\end{align}
where we have used Eqs.~(\ref{xHsingle}) and (\ref{alpha1single}) and $t_0=1/\left( 2H^\ini \right)$ for the last equality.
Now let us take a look at the integral: it is integrated from the initial
time to some later time after the curvaton field decays. Since we are free to choose
the initial time and the integrand is suppressed at later times after the curvaton
decay, we can take the range of integration from zero to infinity without loss of
generality. Then the integral becomes just $\Gamma(3/2) = \sqrt{\pi}/2$. Hence,
\begin{equation}
\frac{\rho_{\gamma 1}}{\rhogzero} = \sqrt{\frac{\pi}{2}}\alpha_1 \approx 1.25331 \alpha_1 \,
.\label{analyticlimitRD}
\end{equation}
Therefore from Eqs.~(\ref{zeta1_out}),~(\ref{f_1_dec}) and~(\ref{analyticlimitRD}),
the final curvature perturbation $\zeta_\gamma^\out$ after the curvaton decays is
\dis{ \zeta_\gamma^\out \approx \frac34\sqrt{\frac{\pi}{2}}\alpha_1\zeta_1^\ini
\approx 0.939986\ \alpha_1\zeta_1^\ini \, , }
which is in good agreement with Ref.~\cite{Gupta:2003jc}.

In the opposite limit ($\alpha_1+\beta_1 \gg 1$), i.e. the curvaton field completely
dominates the energy density of the universe before it decays, mostly the region of
integration is the matter dominated epoch, thus $a \propto t^{2/3}$. We can find the
time of the transition from radiation dominated to curvaton dominated era
($\Omega_{\gamma 0}=\Omega_1$) from sudden decay approximation, \dis{ t_{tr}\approx
\frac{1}{2H^\ini}\left(\frac{\Omega_{\gamma 0}^\ini}{\Omega_1^\ini} \right)^2. }
Using this , the integral becomes
\dis{
\frac{\rho_{\gamma 1}}{\rhogzero}
& \approx \alpha_1 x_H^{1/2}
\left[ \int_{x_0}^{x_{tr}} \left(\frac{t}{t_0}\right)^{1/2} e^{-u_1}du_1
+  \left(\frac{t_{tr}}{t_0}\right)^{1/2}\int_{x_{tr}}^{x_1}\left(\frac{t}{t_{tr}}\right)^{2/3} e^{-u_1}du_1\right],
}
where $x_{tr}=\Gamma t_{tr}$. Ignoring the contribution from the transient radiation
dominated era, we find
\begin{align}\label{analyticlimitMD}
\frac{\rho_{\gamma 1}}{\rhogzero}
& \approx \alpha_1 x_H^{1/2} \left(\frac{t_{tr}}{t_0}\right)^{1/2} \int_{x_0}^{x_1}\left(\frac{t}{t_{tr}}\right)^{2/3} e^{-u_1}du_1
\nonumber\\
&\approx 2^{2/3}\alpha_1\left[ \frac{\Gamma^{(1)}}{H^\ini}\left(\frac{\Omega_{\gamma 0}^\ini}{\Omega_1^\ini}\right)^2 \right]^{-1/6}
\int_0^\infty u_1^{2/3}e^{-u_1} du_1
\nonumber\\
& \approx 2^{2/3}\Gamma\left(\frac{5}{3}\right)
\left(\alpha_1+\beta_1 \right)^{1/3} \alpha_1
\nonumber\\
& \approx 1.43302 \left(\alpha_1+\beta_1 \right)^{1/3} \alpha_1 \, ,
\end{align}
where we assume that initially radiation dominates the universe so that $x_H \approx
\Gamma^{(1)}/H^\ini$ and $t_0 \approx 1/(2H^\ini)$.

For the final matter curvature perturbation,
assuming that initially there is no matter component,
it is independent of the curvaton domination and is simply given from
Eqs.~(\ref{zeta_m_out}) and (\ref{s_i}) by
\begin{equation}
\zeta_m^\out = \zeta_1^\ini \, ,
\end{equation}
i.e. it is just the same as the initial curvature perturbation in the curvaton, as
shown in Ref.~\cite{Gupta:2003jc}.

\subsection{Two curvatons}

\begin{figure}[t!]
  \begin{center}
      \epsfig{file=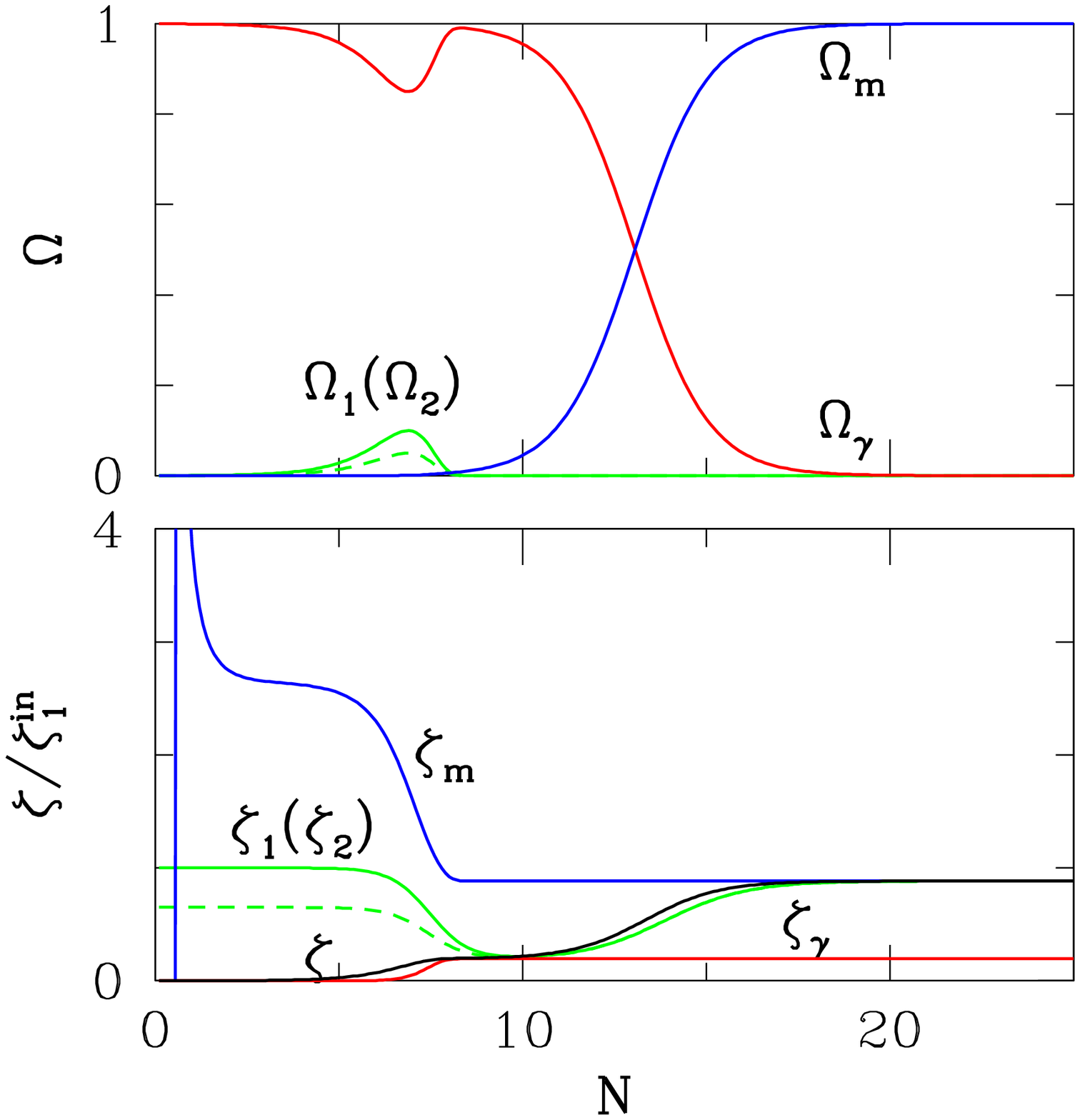, angle = 0, width = 5cm}%
      \epsfig{file=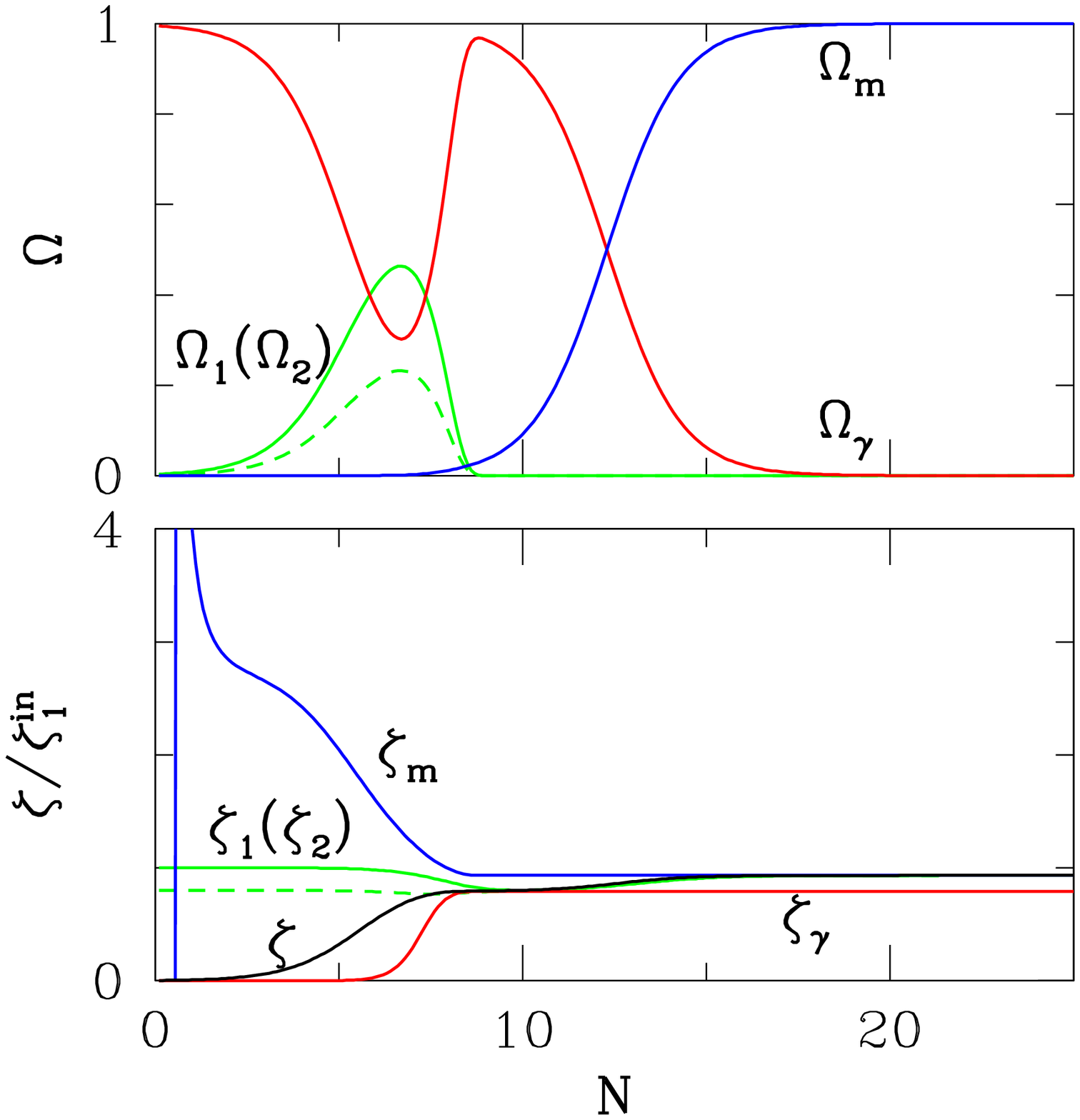, angle = 0, width = 5cm}%
      \epsfig{file=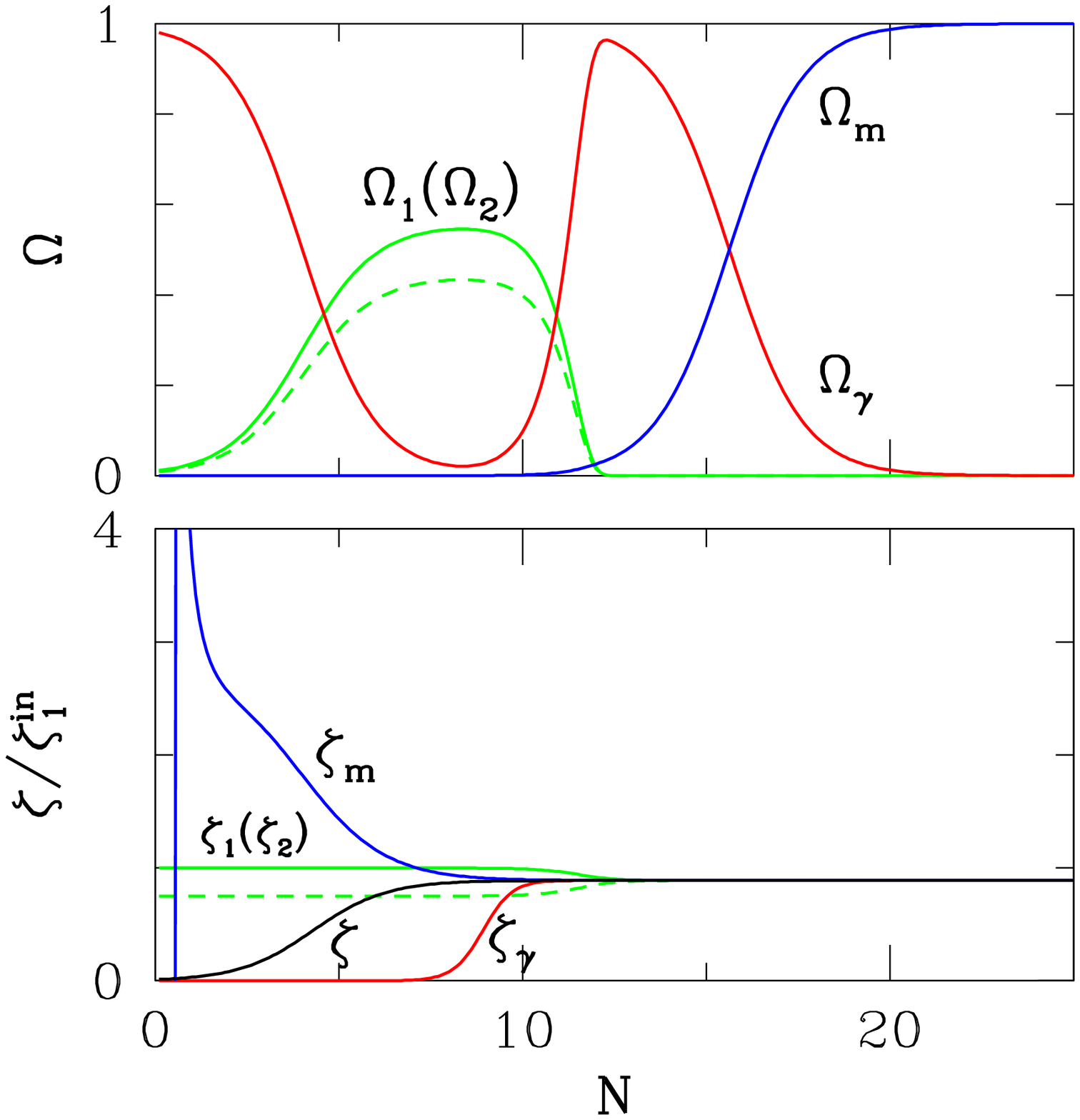, angle = 0, width = 5cm}%
  \end{center}
\caption{The evolution of density parameters (upper row) and curvature perturbations
(lower row) for three cases of two curvaton decays: as shown, the energy densities
of the two curvaton fields are sub-dominant (left panel), dominant (right panel) and
comparable (middle panel) to the radiation energy density. The details are given in
Table~\ref{2curvatontable}.}%
\label{2curvatongraph}
\end{figure}
\begin{table}[h]
\begin{center}
\begin{tabular}{c|c||*{2}{l|}l}
    \multicolumn{2}{c||}{} & left panel &  middle panel & right panel
    \\
    \hline\hline
    \multicolumn{2}{c||}{$\zeta_2^{\ini}/\zeta_1^\ini$} & 0.65 & 0.8 & 0.75
    \\
    \hline
    \multicolumn{2}{c||}{$\Gamma_\gamma^{(1)}/H^\ini$} & $10^{-6}$ & $10^{-6}$ & $10^{-8}$
    \\
    \hline
    \multicolumn{2}{c||}{$\Gamma_\gamma^{(2)}/H^\ini$} & $10^{-6}$ & $10^{-6}$ & $10^{-8}$
    \\
    \hline
    \multicolumn{2}{c||}{$\Gamma_m^{(1)}/H^\ini$} & $10^{-8}$ & $10^{-8}$ & $10^{-10}$
    \\
    \hline
    \multicolumn{2}{c||}{$\Gamma_m^{(2)}/H^\ini$} & $10^{-8}$ & $10^{-8}$ & $10^{-10}$
    \\
    \hline
    \multicolumn{2}{c||}{$\Omega_1^\ini$} & $10^{-3.7}$ & $10^{-2.5}$ & $10^{-2.0}$
    \\
    \hline
    \multicolumn{2}{c||}{$\Omega_2^\ini$} & $10^{-4.0}$ & $10^{-2.8}$ & $10^{-2.1}$
    \\
    \hline\hline
    $r_1$ & analytic approx.& 0.151008 & 0.582546 & 0.556639
    \\
    \cline{2-5}
    & analytic limit & 0.144650 & - & 0.575484
    \\
    \hline
    $r_2$ & analytic approx.& 0.0756833 & 0.291964 & 0.442155
    \\
    \cline{2-5}
    & analytic limit & 0.0724969 & - & 0.423350
    \\
    \hline
    $\zeta_\gamma^\out/\zeta_1^\ini$ & analytic approx.& 0.200202 & 0.816117 & 0.888255
    \\
    \cline{2-5}
    & analytic limit & 0.191773 & - & 0.892997
    \\
    \cline{2-5}
    & numerical & 0.195615 & 0.792049 & 0.887648
    \\
    \hline
    $s_1$ & analytic & 0.666139 & 0.666139 & 0.557312
    \\
    \hline
    $s_2$ & analytic & 0.333861 & 0.333861 & 0.442688
    \\
    \hline
    $\zeta_m^\out/\zeta_1^\ini$ & analytic & 0.883149 & 0.933227 & 0.889327
    \\
    \cline{2-5}
    & numerical & 0.883149 & 0.933228 & 0.889328
\end{tabular}
\end{center}
\caption{The analytic and numerical results for the cases shown in
Fig.~\ref{2curvatongraph}. In the upper half of the table, we give the initial
values used in the calculation and in the lower half we compare the analytic
estimates with the numerical results. For analytic approximation we first solved
Eq.~(\ref{modifiedFriedmann}) to find the density ratio Eq.~(\ref{rhogammaratio})
and then used Eqs.~(\ref{f_i_dec}),~(\ref{zeta_gamma_out}) and~(\ref{r_i}). For
analytic limit we use Eqs.~(\ref{analyticlimitRD}) and~(\ref{analyticlimitMD}) to
find the radiation ratio in the both limits where the curvaton fields remain
sub-dominant / dominant. For final matter curvature perturbation we have used
Eqs.~(\ref{zeta_m_out}) and (\ref{s_i}) for analytic estimate. Note that in the
middle panel we did not use any of the analytic limits, since in this case the
curvatons fields occupy an amount of energy density comparable to radiation and does
not correspond to any of the limiting cases.}%
\label{2curvatontable}
\end{table}

In this section we consider the next simplest case where there are two curvaton
fields decaying into both radiation and matter. If we assume again that the initial
curvature perturbation in radiation is negligible, the final curvature perturbation
in radiation is, from Eq.~(\ref{zeta_gamma_out}),
\begin{align}
\zeta_\gamma^\out & = r_1\zeta_1^\ini + r_2\zeta_2^\ini
\nonumber\\
& = \left( 1 - f_2^\dec \right) f_1^\dec \zeta_1^\ini + f_2^\dec \zeta_2^\ini \, ,
\end{align}
where
\begin{align}
f_1^\dec & = \frac{3\rho_{\gamma 1}/\rhogzero}{4 + 3\rho_{\gamma 1}/\rhogzero} \, ,
\nonumber\\
f_2^\dec & = \frac{3\rho_{\gamma 2}/\rhogzero}{4 \left( 1 + \rho_{\gamma
1}/\rhogzero \right) + 3\rho_{\gamma 2}/\rhogzero} \, .
\end{align}
The final curvature perturbation in matter is given by, from Eq.~(\ref{zeta_m_out}),
\begin{align}
\zeta_m^\out & = s_1\zeta_1^\ini + s_2\zeta_2^\ini
\nonumber\\
& = \frac{\beta_1}{\beta_1 +\beta_2}\zeta_1^\ini
+ \frac{\beta_2}{\beta_1+\beta_2} \zeta_2^\ini \, ,
\end{align}
Therefore the final isocurvature perturbation between radiation and matter is now
completely determined from Eq.~(\ref{finaliso}).
In Fig.~\ref{2curvatongraph},
we show some examples where two curvaton fields decay
into radiation and matter.

 If the energy density of the curvaton fields
remains sub-dominant throughout the evolution of the universe,
which would be guaranteed by the conditions
\dis{
\alpha_1+\beta_1 \ll1, \quad \alpha_2+\beta_2\ll1,
}
 then
\begin{align}
f_1^\dec & \approx c_{R}\alpha_1 \, ,
\nonumber\\
f_2^\dec & \approx c_{R}\alpha_2 \, .
\end{align}
where $c_{R}=3\sqrt{\pi/2}/4\approx0.939986$.

Now it is clear that the transfer coefficients are proportional to the initial
density parameter of the corresponding curvaton fields. Thus with sub-dominant
curvatons the isocurvature perturbation is
\begin{align}
\mathcal{S}_{m\gamma}^\out & = 3\sum_i (s_i - r_i) \zeta_i^\ini
\nonumber\\
&\approx3\left[
\left(\frac{\beta_1}{\beta_1+\beta_2} - c_R\alpha_1\right)\zeta_1^\ini
+\left(\frac{\beta_2}{\beta_1+\beta_2} - c_R\alpha_1\right)\zeta_2^\ini
\right] \, .
\end{align}

For the curvaton dominated case before they decay,
we can take similar steps as radiation dominated one.
For example in the case of the right panel of Fig.~\ref{2curvatongraph},
the transfer coefficients of the matter curvature perturbation $s_i$ are
\dis{
s_{1(2)}=\frac{\alpha_{1(2)}}{\alpha_1+\alpha_2}\,,
}
where we have used $\Gamma_m^{(1)}/\Gamma^{(1)}=\Gamma_m^{(2)}/\Gamma^{(2)}$.
Since the two curvatons dominate at the same epoch, we can use the same
normalisation for $y$-function, thus $f_1$ and $f_2$ are easily approximated as
\dis{
f_1\approx 1\,, \qquad f_2 \approx \frac{\alpha_1}{\alpha_1+\alpha_2}=\frac{\Omega_1}{\Omega_1+\Omega_2}\,,
}
where in the last equality we have used $\Gamma_\gamma^{(1)}/\Gamma^{(1)}=\Gamma_\gamma^{(2)}/\Gamma^{(2)}$.
The transfer coefficients of the radiation curvature perturbation $r_i$ are
\dis{
r_{1(2)}=\frac{\alpha_{1(2)}}{\alpha_1+\alpha_2}\,.
}
The isocurvature perturbation hence almost vanishes, which is shown in the
right panel of Fig.~\ref{2curvatongraph}.

\subsection{Multiple curvatons}

\begin{figure}[t!]
\begin{center}
\epsfig{file=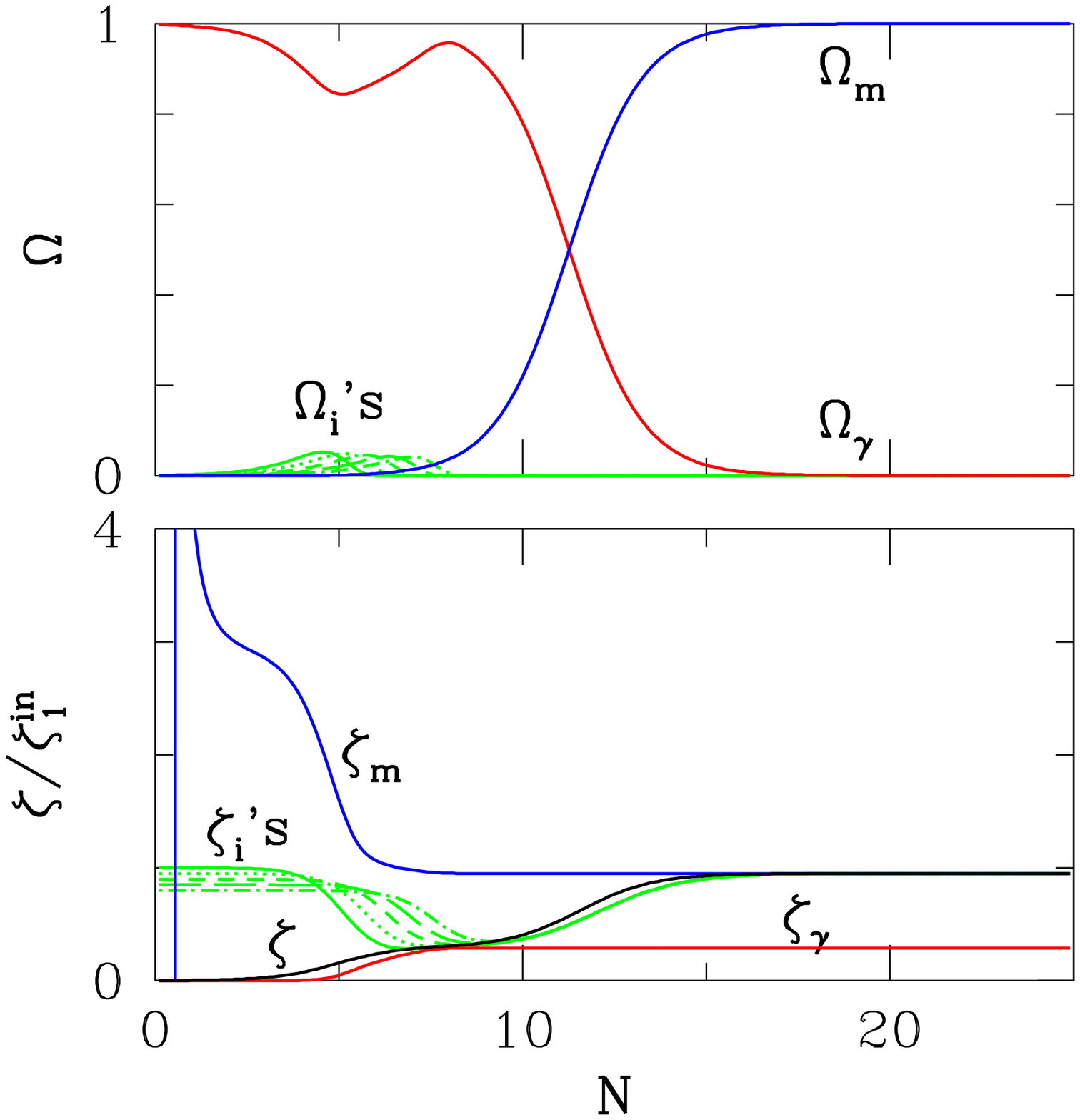, angle = 0, width = 5cm}%
\epsfig{file=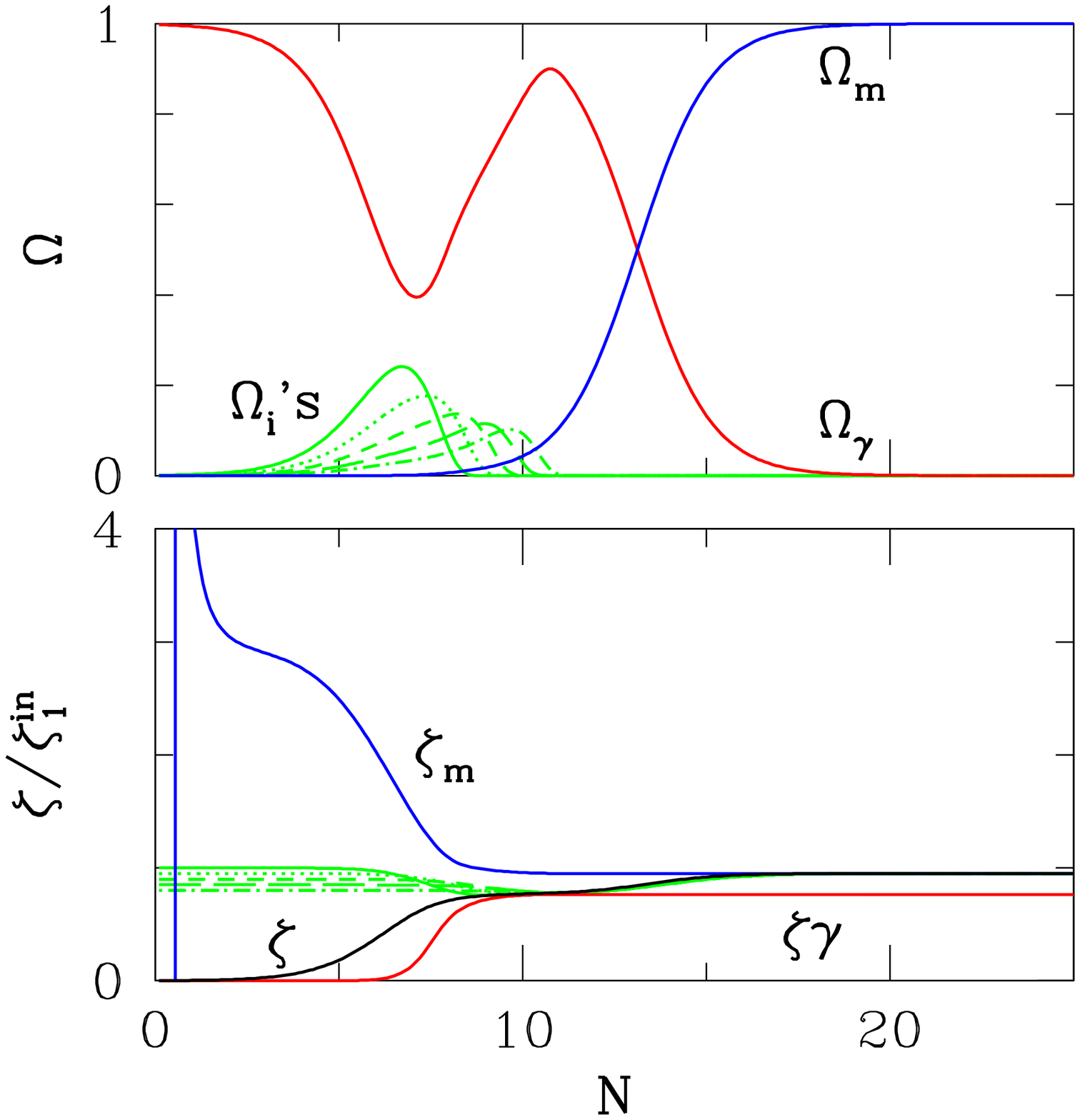, angle = 0, width = 5cm}%
\epsfig{file=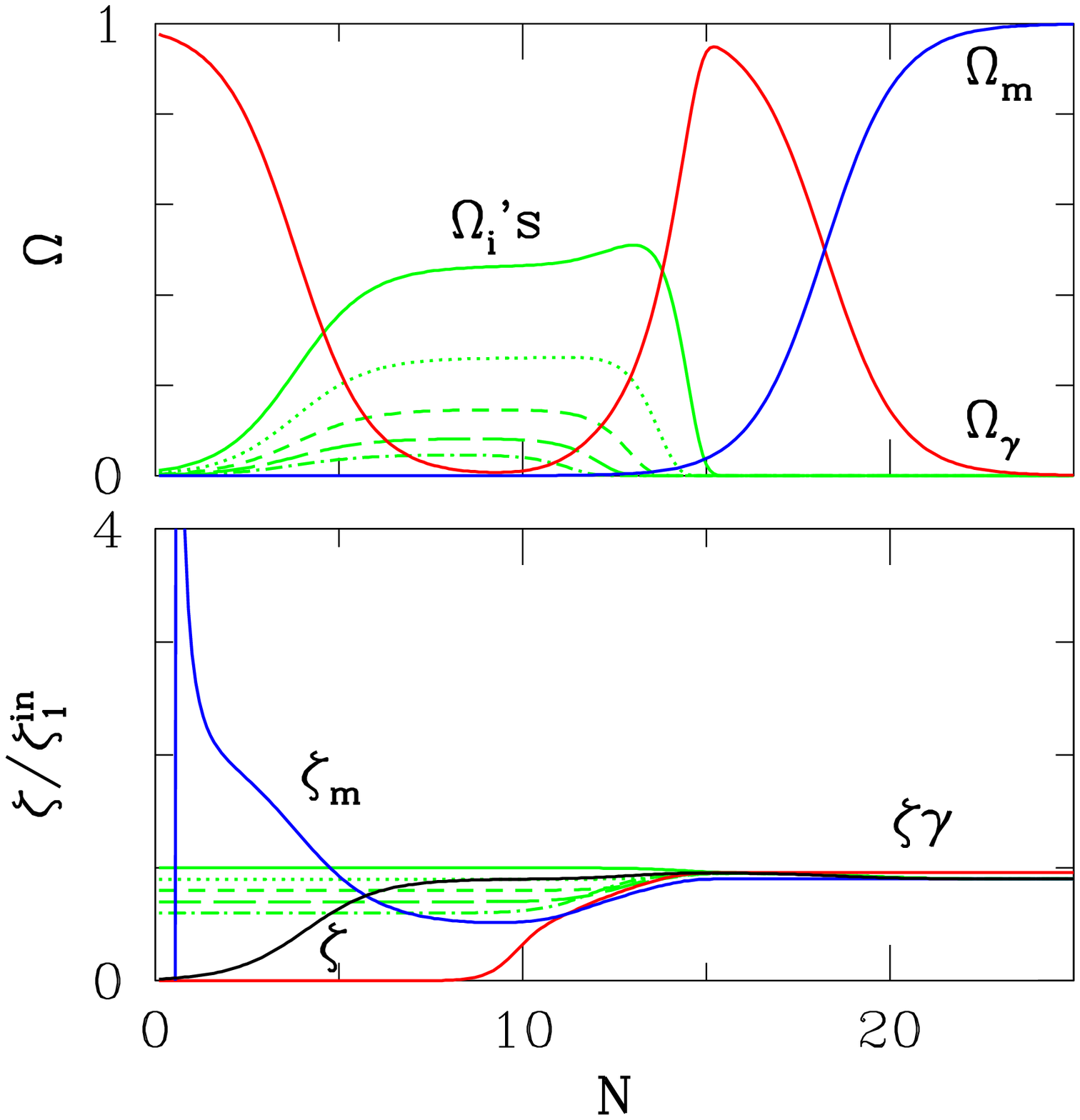, angle = 0, width = 5cm}%
\end{center}
\caption{The same as Fig.~\ref{2curvatongraph} but with five curvaton fields. The
details of the parameters are given in Table~\ref{multitable}. } \label{multigraph}
\end{figure}

\begin{table}[h]
\begin{adjustwidth}{-4em}{-4em}
\begin{center}
\begin{tabular}{c|c||l|l|l}
    \multicolumn{2}{c||}{} & left panel  & middle panel & right panel
    \\
    \hline\hline
    \multicolumn{2}{c||}{$\zeta_i^{\ini}/\zeta_1^\ini$}
&{\small $(1.0,0.95,0.9,0.85,0.8)$} &{\small $(1.0,0.95,0.9,0.85,0.8)$} &{\small
$(0.6,0.7,0.8,0.9,1.0)$}
    \\
    \hline
    \multicolumn{2}{c||}{{\small$\log_{10}\left(\Gamma_\gamma^{(i)}/H^\ini\right)$}}
&{\small $(-4,-4.5,-5,-5.5,-6)$} &{\small $(-6,-6.5,-7,-7.5,-8)$}
&{\small $(-8,-8.5,-9,-9.5,-10)$} \\
    \hline
    \multicolumn{2}{c||}{$\log_{10}\left(\Gamma_m^{(i)}/H^\ini\right)$}
&{\small $(-6,-6.5,-7,-7.5,-8)$} &{\small $(-8,-8.5,-9,-9.5,-10)$ }
&{\small $(-10,-10.5,-11,-11.5,-12)$}   \\
    \hline
    \multicolumn{2}{c||}{$\log_{10}\left(\Omega_i^\ini\right)$}
&{\footnotesize $(-3,-3.25,-3.5,-3.75,-4)$} &{\footnotesize
$(-3,-3.25,-3.5,-3.75,-4)$} &{\footnotesize $(-3,-2.75,-2.5,-2.25,-2)$}
    \\
    \hline\hline
    $r_1$ & analytic approx.& 0.0656799 & 0.187597 & 0.00900859
    \\
    \cline{2-5}
    & analytic limit & 0.0642904 & - & 0.00406052
    \\
    \hline
    $r_2$ & analytic approx.& 0.0650727 & 0.172502 & 0.0256607
    \\
    \cline{2-5}
    & analytic limit & 0.0625240 & - & 0.0142846
    \\
    \hline
    $r_3$ & analytic approx.& 0.0646559 & 0.166930 & 0.0768897
    \\
    \cline{2-5}
    & analytic limit & 0.0609761 & - & 0.0527442
    \\
    \hline
    $r_4$ & analytic approx.& 0.0642456 & 0.163447 & 0.228450
    \\
    \cline{2-5}
    & analytic limit & 0.0596027 & - & 0.196372
    \\
    \hline
    $r_5$ & analytic approx.&0.0637326 & 0.160106 & 0.659821
    \\
    \cline{2-5}
    & analytic limit & 0.0583711 & - & 0.732362
    \\
    \hline
    {\small$\zeta_\gamma^\out/\zeta_1^\ini$} & analytic approx. & 0.291284 & 0.768726 &  0.950306
    \\
    \cline{2-5}
    & analytic limit & 0.275926 & - & 0.963727
    \\
    \cline{2-5}
    & numerical & 0.291515 & 0.765150 & 0.956406
    \\
    \hline
    {\small$\zeta_m^\out/\zeta_1^\ini$} & analytic & 0.950652 & 0.950652 & 0.901304
    \\
    \cline{2-5}
    & numerical & 0.950670  & 0.950652 & 0.901304
\end{tabular}
\end{center}
\end{adjustwidth}
\caption{The analytic and numerical results of in Fig.~\ref{multigraph}. As in
Table~\ref{2curvatontable}, we show the initial parameters in the upper half.}%
\label{multitable}
\end{table}

\begin{figure}[ht]
\begin{center}
\epsfig{file=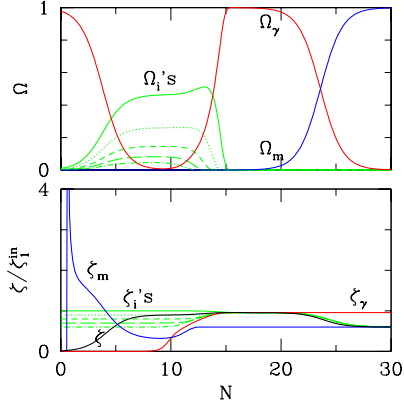, angle = 0, width = 6cm}%
\end{center}
\caption{All the parameters are the same as the right panel of Fig.~\ref{multigraph}
except the branching ratios to matter: here, $\Gamma_i^{(m)}/H^\ini$ is given by
$10^{-19}$, $10^{-17.5}$, $10^{-15}$, $10^{-13.5}$ and $10^{-11}$ for each curvaton,
respectively. In this case, we have $\zeta_\gamma^\out/\zeta_1^\ini = 0.956524$
($\zeta_\gamma^\out/\zeta_1^\ini = 0.950257$ with analytic approx.)  and $\zeta_m^\out/\zeta_1^\ini = 0.602384$,
making $\mathcal{S}_{m\gamma}^\out$ not
negligible. }%
\label{isograph}
\end{figure}

Now we consider more general case where there exist a number of curvaton fields
decaying into radiation and matter. It is straightforward to extract the final
curvature perturbations either numerically by solving
Eqs.~(\ref{Oieq})--(\ref{friedmannOg}) and (\ref{zieq})--(\ref{zeq}), or analytically by
using Eqs~(\ref{finalzeta}) with Eqs.~(\ref{zeta_m_out})
and~(\ref{zeta_gamma_out}). Indeed, as shown in Table~\ref{multitable}, analytic
estimates give good approximations to the full numerical result within an error of
0.7\% (5\%) with analytic approximation (analytic limit). However the evolution of each perturbation could be
quite non-trivial, as shown in Fig.~\ref{multigraph} where we have plotted several
cases with five curvaton fields. We can read the followings:

\begin{itemize}

\item{The evolution of the total curvature perturbation $\zeta$ depends, not surprisingly,
on which component dominates the energy density of the universe. During the curvaton
fields dominates the energy density before they decay, $\zeta$ is the average of
$\zeta_i$'s and constantly evolving during this epoch, since the curvatons are
decaying into radiation and matter. This is clearly seen in the right panel of
Fig.~\ref{multigraph}. After all the curvatons decay, $\zeta$ follows $\zeta_\gamma$
when radiation dominates before matter begins to dominate, and $\zeta = \zeta_m$
afterwards, as shown in Eq.~(\ref{finalzeta}).}

\item{$\zeta_\gamma$ and $\zeta_m$ evolve only during the curvaton fields decay and
remain constant after curvaton fields decay since, as mentioned before, there is no
energy transfer between radiation and matter. Especially, since matter is assumed to
be produced purely due to the decay of the curvatons, $\zeta_m$ is greatly affected
no matter the curvaton fields dominate the energy density before decay or not, e.g.
in the left panel of Fig.~\ref{multigraph} where the curvatons never contribute
significantly, their impact on $\zeta_m$ is large: when $\rho_{m0} = 0$, $\zeta_m$
is just a weighted sum of the initial curvature perturbations of the curvatons and
the weight $s_i$ is basically the ratio of the corresponding curvaton energy density
multiplied by the branching ratio to matter to the total curvaton energy density
responsible to matter density, as shown in Eq.~(\ref{s_i}). For $\zeta_\gamma$, it
is noticeable that $\zeta_\gamma$ becomes significant only when the curvaton fields
occupy significant fraction of total energy density before they decay, as can be
compared between different columns of Fig.~\ref{multigraph}. This is because
practically the radiation is completely generated by the decay of curvaton fields,
making the pre-existing radiation irrelevant.}

\item{From the discussion above, one may tempted to conclude that there will be
negligible isocurvature perturbation between matter and radiation if the curvatons
dominate before they decay, because they are both generated due to the decay of the
curvaton fields. This is not true when there are a number of curvaton fields: the
final isocurvature perturbation is dependent on the background parameters such as
curvaton densities and decay rates. For example, in Fig.~\ref{isograph}, the
branching ratio to matter of the curvaton $\sigma_5$ which has the largest energy
density is extremely small, i.e.
\begin{equation}
\frac{\Gamma_m^{(5)}}{\sqrt{\Gamma^{(5)}}} \approx 10^{-16} \, .
\end{equation}
Thus, although $\zeta_m$ receives contribution from the decay product of the
curvaton with large energy density and this gives a rise of $\zeta_m$, this rise is
never enough to catch up $\zeta_\gamma$ to make $\mathcal{S}_{m\gamma}^\out$
vanishing if the branching ratio is very small as in this case. This is reminiscent
of multi-field inflation: in multi-field inflation, there is no unique prediction on
the isocurvature perturbation produced during inflation. The detail depends on the
inflaton trajectory in the field space. Likewise, generally we can hardly make any
definite prediction on the isocurvature perturbation without the detail.}

\end{itemize}

\section{Conclusions}
\label{conclusions}

In this paper, we have studied the evolution of the universe
which contains a number of non-interacting scalar particles
(the ``curvatons'' $\sigma_i$) decaying into radiation ($\gamma$)
and pressureless matter ($m$) after inflation.
We first have written the evolution equations of the background densities
of the components $\rho_i$, $\rho_\gamma$ and $\rho_m$
which compose the universe and of the curvature perturbations of corresponding
component $\zeta_i$ $\zeta_\gamma$ and $\zeta_m$ on flat hypersurfaces,
Eqs.~(\ref{Oieq})--(\ref{friedmannOg}) and (\ref{zieq})--(\ref{zeq}).
These equations can be numerically solved and give the resulting curvature
perturbations of the components,
as well as the total curvature perturbation $\zeta$
given by Eq.~(\ref{zeta_overall}).

Using the sudden decay approximation,
we have obtained analytic estimates of the final radiation and
matter curvature perturbations $\zeta_\gamma^\out$ and
$\zeta_m^\out$ which are in good agreement with full numerical results.
With the composite densities $\widetilde{\rho}_{\gamma i}$ and $\widetilde{\rho}_{mi}$,
given by Eq.~(\ref{compositedensity}), we can relate $\zeta_\gamma^\out$ and
$\zeta_m^\out$ to the initial curvature perturbations associated with the curvatons
$\zeta_i^\ini$: the curvature perturbation $\widetilde{\zeta}_{mi}$ is conserved and
hence the final matter curvature perturbation $\zeta_m^\out$ has very simple
relation to $\zeta_i^\ini$ Eq.~(\ref{zeta_m_out}) with the transfer coefficient
$s_i$ given by Eq.~(\ref{s_i}). Meanwhile, $\widetilde{\zeta}_{\gamma i}$ is not
constant on large scales since the equation of state of $\widetilde{\rho}_{\gamma i}$ is
not unique. Nevertheless, we can find that $\zeta_\gamma^\out$ is written in terms
of $\zeta_i^\ini$ as Eq.~(\ref{zeta_gamma_out}), with the transfer coefficient $r_i$
given by Eq.~(\ref{r_i}). $r_i$ is determined once the ratio $\rho_{\gamma
i}/\rho_{\gamma 0}$ is found, and we have found a general and model independent
result Eq.~(\ref{rhogammaratio}). This might be also useful to investigate
non-Gaussianity of the primordial curvature perturbation in the multi curvaton
scenario~\cite{Sasaki:2006kq}.

We have applied our results to several different cases.
The analytic estimates give good enough fits to the full numerical results,
within an error of $\mathcal{O}(0.1)$\%.
When the curvatons dominate the energy density of the universe
before they decay, the final radiation curvature perturbation
$\zeta_\gamma^\out$ is significantly affected
by the curvature perturbations of the curvatons $\zeta_i$, since practically radiation is
generated by the decay of the curvaton fields and the pre-existing radiation
is irrelevant.
More importantly, the isocurvature perturbation between matter and radiation
given by Eq.~(\ref{finaliso}) depends on the detailed decay rate of
the curvatons: for example, in the right panel of Fig.~\ref{2curvatongraph},
$\zeta_\gamma^\out$ and $\zeta_m^\out$ are of almost the same amplitudes and
thus isocurvature perturbation is highly suppressed.
However, as shown in Fig.~\ref{isograph},
when the branching ratios to matter are different for different curvatons, we may have significant isocurvature perturbation depending on the initial
values of the background quantities.
We can determine $\mathcal{S}_{m\gamma}^\out$ which may be detected in
the CMB observations only when we have detailed information
on the curvaton fields.

\acknowledgments We thank the organisers of the workshop ``Dark Side of the Universe
2006'' where this work was initiated. KYC acknowledges support from PPARC, and JG is
grateful to Misao Sasaki and the Yukawa Institute for Theoretical Physics where the
early stage of this work was carried out, and Donghui Jeong for useful
conversations.

\end{document}